\documentclass[prd,twocolumn,aps,showpacs,nofootinbib,nobibnotes,
superscriptaddress]{revtex4}
\usepackage{epsfig}
\usepackage{amssymb}
\usepackage{graphics}
\usepackage{bm}

\newcommand{\beq}{\begin{eqnarray}}
\newcommand{\eeq}{\end{eqnarray}}
\begin{document}
\title{Gravitational Radiation Generated by Cosmological Phase Transition
Magnetic Fields}

\author{Tina Kahniashvili}
\affiliation{McWilliams Center for Cosmology and Department of
Physics, Carnegie-Mellon University, Pittsburgh, PA 15213 USA}
\affiliation{Department of Physics, Laurentian University,
Sudbury, ON P3E 2C6, Canada} \affiliation{National Astrophysical
Observatory, Ilia Chavchavadze State University, Tbilisi,
GE-0160, Georgia}

\author{Leonard Kisslinger}
\affiliation{McWilliams Center for Cosmology and Department of
Physics, Carnegie-Mellon University, Pittsburgh, PA 15213 USA}

\author{Trevor Stevens}
\affiliation{Department of Physics, West Virginia Wesleyan
College, Buckhannon, WV 26201 USA}

\date{\today}
\begin{abstract}We study gravitational waves generated
by the cosmological magnetic fields induced via bubble collisions
during the electroweak (EW) and QCD phase transitions. The
magnetic field generation mechanisms considered here are based on
the use of the fundamental EW minimal supersymmetric (MSSM) and
QCD Lagrangians. The gravitational waves spectrum is computed
using a magnetohydrodynamic (MHD) turbulence model. We find that
gravitational wave spectrum amplitude generated by the EW phase
transition peaks at frequency approximately 1-2 mHz, and is of
the  order of $10^{-20}-10^{-21}$; thus this signal is possibly
detectable by Laser Interferometer Space Antenna (LISA). The
gravitational waves generated during the QCD phase transition,
however, are outside the LISA sensitivity bands.
\end{abstract}


\maketitle

\section{Introduction}
There are various mechanisms that might generate gravitational
waves in the early Universe. For reviews
see~\cite{m00,source,hogan}. A well-known one is the generation of
gravitational waves during the cosmological EW or QCD phase
transitions~\cite{EWPT,nicolis,kos1,gs06}. These mechanisms
include bubble wall motions and collisions if the phase
transition is first order~\cite{bubbles}, as well as cosmological
magnetic fields and hydrodynamical or MHD turbulence
~\cite{magnet,kmk,cd,kgr,kkgm08,kcgmr09,Caprini:2009pr}.

    A brief overview of our present work is:

\begin{itemize}

\item We use basic EW and QCD Lagrangians to derive the magnetic
fields created in the plasma during the EWPT and the QCD phase
transitions.

\item In the framework of the standard MHD theory the magnetic field
produces turbulence in the plasma. The value of the
Alfv$\acute{e}$n velocity, $v_A$, is a key parameter when
considering the generation of gravitational waves.

\item From the parameters found via our theory of the EW and QCD
phase transitions, we use the formalism of the gravitational waves
generation by MHD turbulence to estimate the peak frequency and
amplitude of the gravitational waves produced during these
cosmological phase transitions..

\end{itemize}

The direct detection of the relic gravitational waves will open
the new prospects to understand the physical processes in the
early Universe \cite{m00}. The main objective of the present
paper is to study if the gravitational waves produced by the
magnetic fields created during cosmological phase transitions
could be detected by the current and/or nearest future missions.
To be observable the gravitational waves signal must satisfy two
conditions: it must be within the observation frequency bands,
and its amplitude should exceed substantially the instrumental
noise (for the stochastic backgrounds signal to noise ratio (SNR)
must be taken to be 5) \cite{m00}. Our present study is to
determine if the gravitational waves produced either by the EW or
the QCD phase transitions might be detectable by LISA, whose
sensitivity reaches maximum at frequencies $1-100$
mHz~\cite{hogan}.

In order to produce a detectable gravitational wave signal the
cosmological phase transition must be first order, so that
bubbles of the new vacuum nucleate within the false
vacuum~\cite{bubbles} at a critical temperature. Otherwise, there
is a crossover transition.  It has been shown that with the
Standard EW model there is no first order phase
transitions~\cite{klrs}, and there is no explanation of
baryogenesis. However, there has been a great deal of activity in
the MSSM extension of the standard EW model~\cite{r90}. In this
case a MSSM having a Stop with a mass similar to the Higgs mass
leads to the first order EW phase transition. This EW MSSM theory
is consistent with baryogenesis~\cite{laine}. On the other hand,
recent lattice QCD calculations have shown~\cite{qcd} that the QCD
phase transition is a first order cosmological phase transition,
with bubble nucleation and collisions.

The EW phase transition is particularly interesting for exploring
possible cosmological magnetic fields since the electromagnetic
field along with the $W^{\pm}$ and $Z$ fields are the gauge
fields of the Standard model. For the QCD phase transitions the
electromagnetic field is included in the Lagrangian through
coupling to quarks. In both cases the magnetic field with large
enough energy density can have different cosmological signatures.
In particular, the big bang nucleosynthesis (BBN) bound on the
magnetic field energy density is: $\rho_B \leq 0.1 \rho_{\rm rad}
$ \cite{BBN}, giving  $\Omega_B = \rho_B/\rho_{\rm cr} \leq 2.4
\times 10^{-6} h_0^{-2}$, where $\Omega_B$ is the  energy density
parameter at the present time, $\rho_{\rm cr}$ is the present
critical energy density (i.e. the total energy density for the
flat Universe), and $h_0$ is the current Hubble parameter in the
units of 100km/sec/Mpc. This corresponds to a limit on the
effective magnetic field comoving amplitude $B^{\rm eff} =
\sqrt{8\pi \rho_B} \leq 8 \times 10^{-7}$ Gauss.\footnote{In what
follows we use the natural units $c={\hbar}=k_B=1$ and MKS system
for the electromagnetic quantities.} Similar limits can be
obtained through the available data of Laser Interferometer
Gravitational Observatory (LIGO), if it is assumed that the
primordial magnetic field generates relic gravitational waves via
its anisotropic stress \cite{wang}. Stronger limits on the
primordial magnetic field are provided by the cosmic microwave
background (CMB) anisotropy data, $B^{\rm eff}  \leq 5 \times
10^{-9}$ Gauss, see Ref. \cite{CMB-limits} for a review and
references therein.

  Several studies have been performed to estimate the
gravitational wave signal from  the first order EW phase
transition in MSSM and nMSSM, see Refs. \cite{models}. Using a
model similar to the tunneling from a false to a true
vacuum~\cite{bubbles} it was argued that gravitational waves
generated by the EW phase transition are possibly
detectable~\cite{gs06}. However, based on a similar model, it was
concluded that the magnitude of the gravitational waves generated
in a MSSM model of the EW phase transition have a lower amplitude
than that required to be detected by LISA, while in a nMSSM model
the signal is larger~\cite{huber}.

In the present work we re-address the gravitational waves
generated by the primordial magnetic fields created during the
cosmological phase transitions. There are several models of phase
transitions resulting of the magnetic field production \cite{pmf}.
For reviews and references, see Refs.\cite{Widrow,vachaspati}.
Our calculations for the EW phase transition are based on previous
studies of the magnetic field generation by nucleation
\cite{hjk05,jkhhs04} and collisions \cite{sjkhhb08,sj09} of the
EW bubbles. In these studies the basic MSSM EW Lagrangian is used.
Similarly, to generate magnetic fields during the QCD bubble
collisions \cite{jck03,kwj05} the basic QCD Lagrangian is
used~\cite{lsk03}. For the QCD phase transitions the bubble walls
are composed mainly of the gluonic field. After the collision of
two bubbles interior gluonic wall is formed, resulting in a
magnetic wall production (due to coupling of quarks within
nucleons to the gluonic wall causing alignment of nucleon magnetic
dipole moments).

  The magnetic fields generated by the cosmological phase transitions
lead to MHD turbulence.  In the present work we make use of the
basic MHD formulation \cite{orsag70} and determine the main
parameter for MHD turbulence -- the Alfv$\acute{e}$n velocity,
$v_A$, for the magnetic fields produced both by the EW and QCD
phase transitions. The MHD turbulence model described in details
in Refs. \cite{kkgm08,kcgmr09} is then used to calculate the
gravitational waves produced by MHD turbulence present during EW
or QCD  phase transitions. In contrast to the previous works
\cite{EWPT,Caprini:2006rd,kkgm08}, we do not parameterize the
gravitational wave signal in terms of certain phase transitions
parameters, but use our solutions based on the fundamental EW and
QCD Lagrangians. On the other hand, our results are model
dependent in the sense that they use the value of the bubble wall
velocity $v_b=1/2$, found in Ref.~\cite{hjk05}. However, the value
of $v_b=1/2$ was derived using the fundamental EW MSSM theory for
bubble nucleation rather than a model (see below).

  The outline of our paper is as follows.
In Sec. II we review the main MHD equations, the origin of
turbulence, and discuss the main assumptions used in our present
work. In Sec. III we discuss the EW and QCD equations used for
deriving the magnetic fields produced by the two cosmological
phase transitions, and give our results. In Sec. IV we study the
gravitational waves generated by these phase transitions driven by
MHD turbulence. In Sec. V we present and discuss our work. In Sec. VI
we give our conclusions.

\section{MHD TURBULENCE MODEL}

  Our objective is to derive the gravitational wave energy density,
$\rho_{\rm GW}$ produced by the magnetic field, ${\bf B}$,
created during  EW and QCD phase transitions. In the first
subsection section we discuss the basic MHD theory leading to
turbulence, and in the second subsection review the MHD turbulence
model used for our cosmological applications.

\subsection{MHD Turbulence and the Alfv$\acute{e}$n Velocity}

 An essential aspect of our model is that
the magnetic field created after bubble collisions couples to the
plasma and creates MHD turbulence, which then generates via the
anisotropic stress the gravitational waves. In other words, the
magnetic energy density $\rho_B$ is transformed  to the
gravitational waves energy density $\rho_{\rm GW}$. Since the
coupling between the magnetic and gravitational waves energy
occurs through the Newton gravitational constant $G$, due to the
small value of $G$ the efficiency of the gravitational wave
production directly from the magnetic field is too small, see also
Refs. \cite{Caprini:2009pr}, but even accounting for the small
efficiency, the gravitational wave signal can be possibly
detectable.

To show coupling between an initial phase transition generated
magnetic field to the plasma we give here the basic MHD equations
for an incompressible, conducting fluid~\cite{orsag70}
\begin{eqnarray}
\left[\frac{\partial}{\partial\eta} + ({\bf v} \cdot \nabla )- \nu
\nabla^2\right] {\bf v} &=& ({\bf b} \cdot \nabla ) {\bf b}-\nabla
p + {\bf f}, \label{mhd1} \\
\left[\frac{\partial}{\partial \eta} -\eta_{\rm re}
\nabla^2\right] {\bf b} &=& -({\bf v}\cdot\nabla ){\bf b}
+ ({\bf b}\cdot \nabla ) {\bf v}, \label{mhd2} \\
               \nabla {\bf v} &=&  \nabla {\bf b}=0 \; \label{mhd3},
\end{eqnarray}
where $\eta$ is the conformal time, ${\bf v}({\bf x},\eta)$ is
the fluid velocity, ${\bf b}({\bf x},\eta) \equiv {\bf B}({\bf
x},\eta)/ \sqrt{4\pi {\rm w}}$ is the normalized magnetic
field,  ${\bf f}({\bf x},\eta)$ is an external force driving the
flow, $\nu$ is the comoving viscosity of the fluid,  $\eta_{\rm re}$ is
the comoving resistivity, and w$=\rho +p$, $\rho$, and $p$ the
enthalpy, energy density, and pressure of the plasma.

 The coupling of the magnetic field to
the plasma (see Eqs. (\ref{mhd1})-(\ref{mhd3}) leads to
Alfv$\acute{e}$n turbulence development with the characteristic
velocity, $v_A$ \cite{B03,son},  \beq \label{vA}
    v_A &=& \frac{B}{\sqrt{4\pi {\rm w}}}\;
= \sqrt{\frac{3\rho_B}{2 \rho}} \;. \eeq Since both $\rho_B$ and
$\rho=\rho_{\rm rad}$ scale as $1/a^4(t)$, (where $a(t)$ is the
cosmic scale factor), if there is no damping of the magnetic
field (additional $\rho_B$ temporal dependence), the value of
$v_A$ is not affected by the expansion of the Universe. $v_A$ is
an essential parameter in the generation of gravitational waves
in the magnetized turbulent model considered here (for details of
the hydro-turbulence model, see Ref.~\cite{kkgm08}). To develop
the turbulence picture it is assumed apriori that the EW or QCD
bubble collisions lead to the vorticity fluctuations, i.e. the
presence of the kinetic turbulence, with a characteristic
velocity (associated with the largest size bubble) $v_0$.
Equipartition between the magnetic and kinetic energy densities
implies $v_A \simeq v_0$.{\footnote{Accounting for the stochastic
nature of the magnetic field, the Alfv\'en velocity is scale
dependent, and if it is not specified, $v_A$ is associated with
the largest size magnetic eddy. On the other hand, $v_A$ can be
expressed in terms of the magnetic field comoving amplitude as,
\begin{equation}
v_A \simeq 4 \times 10^{-4} \left( \frac{B}{10^{-9}{\rm
Gauss}}\right)\left(\frac{g_\star}{100}\right)^{-1/6}.
\label{va1}\end{equation} Here we used Eqs. (\ref{vA}) and
(\ref{rho}). The MHD turbulence description presumes that the
magnetic turbulent energy density is saturated when the Alfv\'en
velocity reaches the kinetic velocity of the plasma, i.e.
$v_A\simeq v_0$.}}  We also define the energy density of the
plasma to be equal of the radiation energy density, $\rho_{\rm
rad}$, and $\rho_{\rm rad}$ at the moment of the phase transition
with temperature $T_\star$, is given by
\begin{equation} \label{rho}
    \rho_{\rm rad}({\rm at}~T_\star)  = \frac{\pi^2}{30}g_* (T_*)^4 \; ,
\end{equation} where  $g_\star$ is the number of relativistic degrees of
freedom at temperature $T_\star$ \cite{kt90}. Using the BBN bound
on the total magnetic energy density $\rho_B < 0.1 \rho_{\rm
rad}$, the Alfv\'en velocity must satisfy $v_A \leq 0.4$.

The Afv\'en velocity $v_A$ as well as the bubble kinetic motion
velocity $v_0$ can be related to the phase transitions
parameters,  $\alpha_{\rm PT}$, the ratio between the latent heat
and the thermal energy, and the efficiency, $\kappa_{\rm PT}$,
which determines what part of the vacuum energy is transferred to
the kinetic energy of the bubble motions as opposed to the
thermal energy. Ref. \cite{nicolis} presents the estimate for the
largest size turbulent bubble velocity, $v_A \simeq v_0 =
\sqrt{\kappa_{\rm PT} \alpha_{\rm PT}/(4/3 + \kappa_{\rm PT}
\alpha_{\rm PT})} $, and thus accounting for $v_A \simeq v_0$,
BBN bound leads $\kappa_{\rm PT} \alpha_{\rm PT} \leq 0.2$.

\subsection{Direct Cascade and Magnetic Energy Density}
After being coupled to the fluid, the magnetic field energy
density injected into the plasma at characteristic comoving
length scale $\lambda_0 $ (which is, of course, inside the
comoving Hubble radius $\lambda_H$ at the moment of the field
generation). The magnetic field energy density due to coupling
with the fluid motions is re-distributed spatially through the
following  regimes \cite{B03}, $k<k_0$ ($k_0=2\pi/\lambda_0$),
$k_0 <k<k_D$ (with $k_D=2\pi/\lambda_D$ the wavenumber
corresponding to the magnetic field damping scale due to the
plasma viscosity) and $k>k_D$. Inside so-called inertial range,
$k_0 <k<k_D$, the selective turbulence decay occurs and the
magnetic energy flows from the large to the small scales
according to the direct cascade Kolmogoroff law \cite{P52},
resulting in the magnetic field spectral energy density $E_M(k)
\propto k^{-5/3}$, while at large scales ($k<k_0$) the free
turbulence decay takes place, leading to the magnetic field
spectrum $E_M(k) \propto k^{\alpha_T}$, with $\alpha_T \geq 4$
\cite{caprini-durrer}. The initial Batchelor spectrum
$\alpha_T=4$ can be transformed via the non-linear processes to
the Kazantzev spectrum, $\alpha_T=3/2$ \cite{axel}. Another
numerical realization of large scale turbulence can be the white
noise spectrum, $\alpha_T=2$, (Saffman) spectrum \cite{hogan1}.
Using the fact that the total energy density of the magnetic
field $E_M =\int_0^\infty dk~E_M(k) $ can not be larger than the
initial magnetic energy density, $\rho_B$, it is straightforward
to obtain the maximal allowed values for $E_M(k)$, and get the
magnetic field limits at large scales \cite{ktr09}.

In our present work MHD turbulence is created by bubble
collisions. It is physically justified to assume that the typical
injection scale of  the magnetic energy is associated with the
phase transition largest bubble sizes, which are given by the
bubble wall velocity $v_b$ and phase transition time-scales
$\beta^{-1}$ for the EW and QCD phase transitions (determined
through the bubble nucleation rate), i.e. $\gamma \equiv
\lambda_0/\lambda_H = v_b (\beta/H_\star)^{-1} \ll 1$
\cite{kos1}. With these assumptions one finds for the Kolmogoroff
power law and the wave numbers\cite{kkgm08}
\begin{equation}
         E_{M}(k,t) =  C_M \varepsilon^{2/3} k^{-5/3}
\end{equation} over the range of wave numbers $k_0 < k < k_D$, where
$k_D = k_0 Re^{3/4}$, $C_M \simeq 1$, $R \gg 1$ is the turbulence
Reynolds number at the temperature $T_\star$, and $\varepsilon =
(2/3)^{3/2} k_0 v_A^3 $ is the comoving magnetic energy
dissipation rate per unit enthalpy.

  We consider only the inertial (direct cascade) range due to following
reasons \cite{B03}:
  i) helicity vanishes (for the
effects of initial kinetic or magnetic helicity, see
Refs.~\cite{kgr}.) in our symmetric treatment of bubble
collisions for the phase transitions; ii) we presume that the
contribution from the large scales ($k<k_0$) into the
gravitational waves signal will not exceed significantly (or even
being smaller) than that which comes from the inertial range, due
to the free decay of non-helical turbulence and the magnetic
energy density small amount presence at large scales (we will
address this issue in the separate work \cite{ktr}). Another
important assumption that we make is equipartition between the
kinetic and magnetic energy densities, which allows us to use the
direct analogy between the hydro and magnetized turbulence.

The direct cascade turbulence is characterized not only by the
spatial structure ($k$-dependence), but it is important to take
into account  the time dependence of the turbulent quantity
correlations. First, we assume that the source lasts enough time
to allow us to consider the developed turbulence, so we do not
include in our calculations the pulse-like source
\cite{Caprini:2009pr}. To compute the direct cascade duration
time we use the fluctuation time decorrelation function $\eta(k)
\simeq \sqrt{2\pi} (k/k_0) ^{2/3} \tau_0^{-1}$ \cite{K64}, where
$\tau_0=l_0/v_0$ is the largest turbulent eddy turn-over time.
Note $l_0=v_b \beta^{-1}$ is the physical length scale of the
largest size eddy (bubble). As a result the time dependence of the
magnetic field two-point correlation function within the inertial
range is given by the function $f(\eta(x),\tau)$ \cite{K64}. The
function $f(\eta(x),\tau)$ is such that it becomes negligibly
small for $\tau \gg 1/\eta(k)$, and from the dimensional analysis
is adopted $f[\eta(k),\tau] = \exp \! \left[- \pi \eta^2(k)
\tau^2/4 \right]$ \cite{K64}. It is clear that the temporal
decorrelation function $\eta(k) \propto (k/k_0)^{2/3} $ reaches
its maximum equal to $ \sqrt{2\pi}/\tau_0 = \sqrt{2\pi} \beta v_A
v_b^{-1}$ within the inertial range for the maximal size eddy,
$k=k_0$, as a consequence, the smallest eddies are decorrelated
first, and the turbulence cascade time is determined by the
largest size eddy turn-over time $\tau_0$.
As a result the comoving (measured today) peak frequency of the
induced gravitational waves is given by $f_{\rm peak} \simeq
v_A/\lambda_0$ (see below).

 The proper consideration of
the temporal decorrelation leads to the fast damping of the
gravitational wave signal amplitude for frequencies $f\gg f_{\rm
peak}$, i.e. larger than that associated with the direct cascade
turbulence induced peak frequency, \cite{kkgm08,kcgmr09}. Several
previous studies, Refs. \cite{kmk,cd}, did not account for the
temporal exponential decorrelation function, and as a result the
shape of the gravitational wave at high frequencies was given by
power law, without having steep exponential damping
\cite{kkgm08}. We also underline that our description does not
apply for any pulse-like sources \cite{Caprini:2009pr}. In the
former case, the characteristic comoving gravitational wave peak
frequency is determined by $1/\lambda_0$ \cite{cd}, so it is
higher than that in the case considered here.

Another consequence of the temporal decorrelation is that the
turbulence cascade time-scale is much shorter than the Hubble
time scale, and thus we are allowed neglect the expansion of the
Universe during the gravitational waves generation process
\cite{ktr}. We account for the expansion of the Universe only when
computing the gravitational wave amplitude $h_C(f)$ measured today
(or the corresponding spectral energy density parameter
$\Omega_{\rm GW}(f)$), as a function of the linear frequency $f$
measured today.\footnote{The amplitude and the energy density of
the gravitational wave is related through \cite{m00},
\begin{equation} h_C(f) = 1.26 \times 10^{-18} \left( \frac{\rm
Hz}{f} \right) \left[ h_0^2 \, \Omega_{\rm GW}(f) \right]^{1/2},
\label{gw-amplitude}
\end{equation}
where $f$ is the linear frequency measured today, and it is given
by $f=(2\pi)^{-1}\omega$, with $\omega=(a_\star/a_0)
\omega_\star$, where $\omega_\star$ is the gravitational wave
frequency at the time of generation. Due to the Universe
expansion the freely propagating gravitational wave amplitude and
frequency are rescalled by a factor \beq \label{a-ratio}
\frac{a_\star}{a_0} &\simeq& 8 \times 10^{-16}
\left(\frac{100\,{\rm GeV}}{T_\star}\right)
\left(\frac{100}{g_\star}\right)^{{1}/{3}} \; , \eeq where
$a_\star$ and $a_0$  are the scale factors at the time of
generation and today, respectively. See Ref.~\cite{m00} for the
definitions and discussions on gravitational wave direct detection
experiments.}

\section{MAGNETIC FIELDS GENERATED BY EW AND QCD PHASE TRANSITIONS}

\subsection{Electroweak Phase Transition}

   First we review the magnetic field created during EW phase transition.
In a suitable MSSM model the EW phase transition is first order,
which results in bubble nucleation and collisions. The MSSM EW
theory for bubble nucleation is developed in Ref. \cite{hjk05}, a
Weinberg-Salam model with all supersymmetric partners integrated
out except the Stop, the partner to top quark, has the form \beq
\label{L}
  {\cal L}^{MSSM} & = & {\cal L}^{1} + {\cal L}^{2}  + {\cal L}^{3}
\nonumber  \\
      && +{\rm leptonic \: and \: quark \: interactions }\\ \nonumber
         {\cal L}^{1} & = & -\frac{1}{4}W^i_{\mu\nu}W^{i\mu\nu}
  -\frac{1}{4} B_{\mu\nu}B^{\mu\nu} \\ \nonumber
 {\cal L}^{2} & = & |(i\partial_{\mu} -\frac{g}{2} \tau \cdot W_\mu
 - \frac{g'}{2}B_\mu)\Phi|^2  -V(\Phi) \nonumber \\
 {\cal L}^{3} &=& |(i\partial_{\mu} -\frac{g_s}{2} \lambda^a C^a_\mu)\Phi_s|^2
    -V_{hs}(\Phi_s,\Phi) \nonumber \, ,
\eeq with \beq \label{wmunu}
   W^i_{\mu\nu} & = & \partial_\mu W^i_\nu - \partial_\nu W^i_\mu
 - g \epsilon_{ijk} W^j_\mu W^k_\nu\\ \nonumber
 B_{\mu\nu} & = & \partial_\mu B_\nu -  \partial_\nu B_\mu \, ,
\eeq where the $W^i$, with i = (1,2), are the $W^+,W^-$ fields,
$C^a_\mu$ is an SU(3) gauge field, ($\Phi$, $\Phi_s$) are the
(Higgs, right-handed Stop fields), $(\tau^i,\lambda^a)$ are the
(SU(2), SU(3)) generators, and the EM and Z fields are defined as
\beq
\label{AZ}
   A^{em}_\mu &=& \frac{1}{\sqrt{g^2 +g^{'2}}}(g'W^3_\mu +g B_\mu) \nonumber \\
   Z_\mu &=& \frac{1}{\sqrt{g^2 +g^{'2}}}(g W^3_\mu -g' B_\mu) \; .
\eeq

The  parameters used here are $g=e/\sin\theta_W = 0.646$ and
$g'=g \; \tan\theta_W =0.343$. The equations of motion were
solved using an SU(2) I-spin ansatz for the gauge fields, and it
was found that the nucleation velocity of the bubble walls,
$v_b=1/2$. This is a very important parameter for our present
work. Using EW MSSM theory, we find $v_b$ to be larger than found
using models for EW bubble walls \cite{moore00}.

   Starting from the Lagrangian given by Eq.~(\ref{L}) for the
MSSM EW phase transition, from which bubbles are formed, to
compute bubble collisions we assume that the Stop is integrated
out, so the ${\cal L}^{3}$ term is not included. The magnetic
field created during the collision of two bubbles was estimated by
Refs.~\cite{sjkhhb08,sj09}. Our model is based on the assumption
that the final magnetic field is created by the final two bubbles
colliding. The time scale is from $10^{-11} \; $ to $ \;
10^{-10}$ seconds, and the critical temperature (EW phase
transition energy scale) is $T_* \simeq  M_H$, with the Higgs mass
$M_H \simeq 110 {\rm \; to\; } 130$ GeV \cite{H}.
  The magnitude of the magnetic field created in this final collision of two
bubbles, $B^{(\rm EW)}$, is found in our new calculations
\cite{sj209}, based on Ref.~\cite{sj09}, but with the colliding
bubbles having larger overlap. That is the bubble collision has
been followed to the stage of overlap of the bubbles such that
the B-field in the Universe just after the EW phase transition has
been determined. The result for the B-field is \beq \label{bewpr}
        B^{(\rm EW) }_\star ({\rm at }~~ T_\star^{({\rm EW} )})  &\simeq&  10
        M_W^2 \nonumber \\
               &=& 6.4 \times 10^{4} M_{W,80}^2 {\rm GeV}^2 \; .
\eeq where $M_{W,80} = {M_W}/{80{\rm GeV}}$ is normalized W-boson
mass. Even though the above estimate is given for two bubbles
collision, we extend our consideration presuming that there is a
continuous creation of the  magnetic field through the bubble
collisions, and the total magnetic energy density released during
the MSSM EW phase transition can be approximated by $\rho_B^{\rm
EW}= (B^{(\rm EW)})^2/8\pi$. In reality of course there are many
bubble collisions,\footnote{It is clear that within the Hubble
radius we have several areas where the magnetic field is
generated. The size of the colliding bubble determines the
correlation length of the magnetic field, as we noted above.}
which leads to the establishment of the stochastic nature of the
resulting magnetic field. The coupling between this initial field
with the fluid results in MHD turbulence \cite{son}.
  Using Eqs. (\ref{vA})-(\ref{rho}) we find that the Alfv\'en velocity is
\begin{equation} \label{valven}
  v_A^{({\rm EW})} \simeq 0.27 M_{W,80}^2 \left(\frac{T_\star}{100{\rm GeV}}
\right)^{-2}\left(\frac{g_\star}{100}\right)^{-1/2}.
\end{equation}
 For the EW phase transition with $g_* \simeq 100$ and $T_* \simeq 100$ GeV,
 one finds that $ v_A^{(\rm EW)} \simeq 0.27$, which is below the BBN bound
discussed above.  On the other hand the value of $v_A \simeq 0.27$ means
that $7.4\%$ of the radiation energy density is in the form of the
magnetic energy density.

\subsection{QCD Phase Transition}

Recent lattice QCD studies have shown that the
QCD phase transition is first order \cite{qcd}.
In Ref.~\cite{lsk03} the magnetic field created by the QCD phase transition
bubble collisions was derived. A gluonic wall is created as two
bubbles collide, and a magnetic wall is formed by the interaction
of the nucleons with the gluonic wall, with electromagnetic
interaction Lagrangian \beq \label{Lint}
 {\cal L}^{\rm int} &=&  -e \bar{\Psi} \gamma^\mu A^{em}_\mu \Psi,
\eeq where $\Psi$ is the nucleon field operator and $A^{em}$ is
the electromagnetic 4-potential. In Ref. \cite{lsk03} it was shown
that the interaction of the quarks in the nucleons with the
gluonic wall alignes  the nucleons magnetic dipole moments,
producing a B-field orthogonal to the gluonic wall.

    Using an instanton model, for the gluonic instanton wall oriented in
the x-y direction one obtains for $B_z \equiv B^{(\rm QCD)}_\star
$ within the wall, of thickness $\zeta$ \beq
 \label{bz}
     B^{(\rm QCD)}_\star ({\rm at} ~ T_\star^{({\rm QCD})})  &\simeq& \frac{1}{\zeta\Lambda_{\rm QCD}} \frac{e}{2 M_n}
\nonumber \\ &\times &
 < \bar{\Psi} \sigma_{21}\gamma_5 \Psi > \; ,
\eeq
where $\Lambda_{\rm QCD}$ is the QCD momentum scale.
A similar form had been derived earlier using a domain wall
model~\cite{fz00}. The value for $B^{(\rm QCD)}$ was found to be
\beq
\label{bw}
   B^{(\rm QCD)}_\star &\simeq& 0.39 \frac{e}{\pi} \Lambda_{\rm QCD}^2 \simeq
          1.5 \times 10^{-3} {\rm GeV^2} \; .
 \eeq
Eqs.~(\ref{vA})-(\ref{rho}) with $g_*$=15, $T_\star$=0.15 GeV give
\beq
\label{vaqcd}
            v_A^{({\rm QCD})} &\simeq& 8.4\times 10^{-3}.
\eeq

\section{GRAVITATIONAL WAVES GENERATED BY MAGNETIC FIELDS}

In this section we derive the strain amplitude measured today,
$h_C(f)$,  of the gravitational waves generated by MHD turbulence
developed during the EW or QCD phase transitions. The
gravitational waves are generated by the transverse and traceless
part, $S_{ij}$, of the stress-energy tensor, $T_{ij}$
\cite{mtw73}. \beq \label{sij}
     S_{ij}({\bf x},t) &=& T_{ij}({\bf x},t) -\frac{1}{3}
\delta_{ij}T_k^k({\bf x},t) \; . \eeq When considering
gravitational waves generated by the magnetic field, the source
$S_{ij}$ is associated with the magnetic anisotropic stress
\cite{magnet}. On the other hand, if the duration of the source
is short enough when comparing with the Hubble time at the moment
of the generation, $H_\star^{-1}$, we can neglect the expansion
of the Universe and the gravitational wave generation is
described by the simplified equation and solution \beq
\label{eq:01} \nabla^2 h_{ij}({\mathbf x}, t)
-\frac{\partial^2}{\partial t^2}
h_{ij}({\mathbf x}, t) &=& -16\pi G S_{ij} ({\mathbf x}, t) \nonumber \\
       h_{ij}({\mathbf x}, t) &=&\int d^3{\bf x}' \frac{S_{ij}({\mathbf x}',t)}
{|{\bf x}'-{\bf x}|} \; . \eeq Here $h_{ij}({\mathbf x}, t)$ is
the metric tensor perturbation which satisfies the following
conditions: $h_{ii}=0$ and $\partial h_{ij}/\partial x^j=0$.

To derive the energy density of the induced gravitational waves,
accounting for the stochastic nature of the magnetic turbulence
source, and as a consequence the stochastic nature of the
gravitational signal, we must compute the autocorrelation function
$\langle
\partial_t h_{ij}({\mathbf x},t) \partial_t h_{ij} ({\mathbf x},t+\tau)
\rangle/ 32\pi G$ which can be expressed through the  two-point
correlation function of the source $R_{ijij}({\bm \xi},
\tau)=\langle S_{ij}({\mathbf x}^\prime, t) S_{ij}({\mathbf
x}^{\prime \prime},t+\tau) \rangle$, where ${\bm \xi}={\bf
x}^{\prime \prime}-{\bf x}^\prime$ \cite{kkgm08}. As the
calculations performed in Refs. \cite{kkgm08,kcgmr09} show the
gravitational energy density, $\rho_{\rm GW}$ at the moment of
generation is given by the duration time $\tau_T$ and the Fourier
transform of $R_{ijij}$ tensor, $H_{ijij}(0,\omega_\star)$,
\beq
 \label{spectrum}
  \rho_{\rm GW}(\omega_\star) &=& 16\pi^3\omega_\star^3 G\,{\rm w}_\star^2
\tau_T H_{ijij}(0,\omega_\star) \; ,
\eeq
where $\omega_\star$ is the
angular frequency measured at the moment of the gravitational
waves generation. To obtain $\rho_{\rm GW}(\omega_\star)$ and the
frequency today, one must account for the gravitational wave
amplitude and frequency rescaling given by Eq. \ref{a-ratio}.

  Since equipartition between kinetic and magnetic energy
densities is maintained during Kolmogoroff turbulence, the total
source for the gravitational waves (from the magnetic and kinetic
turbulence) is simply two times the kinetic turbulence source.
Using the assumptions made above, the source $H_{ijij}$ tensor is
given by \cite{kkgm08,kcgmr09},
\begin{eqnarray}
 H_{ijij}(0,\omega_\star) & = &  \frac{7 C_M^2
{\bar \varepsilon}}{6 \pi^{3/2} }  \int_{{\bar k}_0}^{{\bar k}_D}
\! \frac{d{\bar k}}{{\bar k}^6} \exp\!\left(
-\frac{\omega_\star^2}{{\bar \varepsilon}^{2/3} {\bar k}^{4/3}} \right)\nonumber \\
&&{\rm erfc} \!\left( -\frac{\omega_\star}{{\bar
\varepsilon}^{1/3} {\bar k}^{2/3}} \right).~  \label{H}
\end{eqnarray}
Here, ${\rm erfc}(x)$ is the complementary error function defined
as $\mbox{erfc}(x) = 1 - \mbox{erf}(x)$, where $\mbox{erf}(x) =
\int_0^x dy \exp(-y^2)$ is the error function. ${\bar\varepsilon}
= (a_0/a_\star) \varepsilon  $ and ${\bar k} = (a_0/a_\star) k_0$
are the physical energy dissipation rate and the physical
wavenumber respectively.  As can be expected, the integral in
Eq.~(\ref{H}) is dominated by the large scale (${\bar k} \simeq
{\bar k}_0$) contribution so, for forward-cascade turbulence, the
peak frequency is~\cite{kkgm08} $ \omega_{\rm max. \star} \simeq
{\bar k}_0 v_0$. It must be noted that the peak frequency is
determined by the time characteristic of turbulence only in the
case when the turbulence duration time is enough long ($\tau_T
\simeq {\rm few} \times \tau_0$  to insure the applicability of
the Proudman argument ~\cite{P52}, which has been used to justify
the use of the Kolmogoroff model. Otherwise, the peak frequency
is determined by the characteristic scales of the pulse-like
source \cite{Caprini:2006rd,Caprini:2009pr}.

 Taking into account the expansion of the Universe
and using Eq.~(\ref{H}) for the $H_{ijij}$ tensor, we find the
gravitational wave amplitude as a function of the linear frequency
measured today, $f = (a_\star/a_0) f_\star$ with $f_\star =
\omega_\star/2\pi$, \beq \label{hctoday} h_C(f) &\simeq & 2
\times 10^{-14} \left(\frac{100\,{\rm GeV}}{T_*} \right)
\left(\frac{100}{g_*}\right)^{1/3} \nonumber
\\ && \times \left[\tau_T \omega_\star H_\star^4
H_{ijij}(0, \omega_\star)\right]^{1/2}. \eeq where $f_H =
\lambda_H^{-1}$  is  the Hubble frequency measured today, $f_H
\simeq 1.6 \times 10^{-5}\,{\rm Hz}\, ({g_*}/{100})^{1/6}
({T_*}/{100\,{\rm GeV}})$ .

\begin{figure}[ht]
\epsfig{file=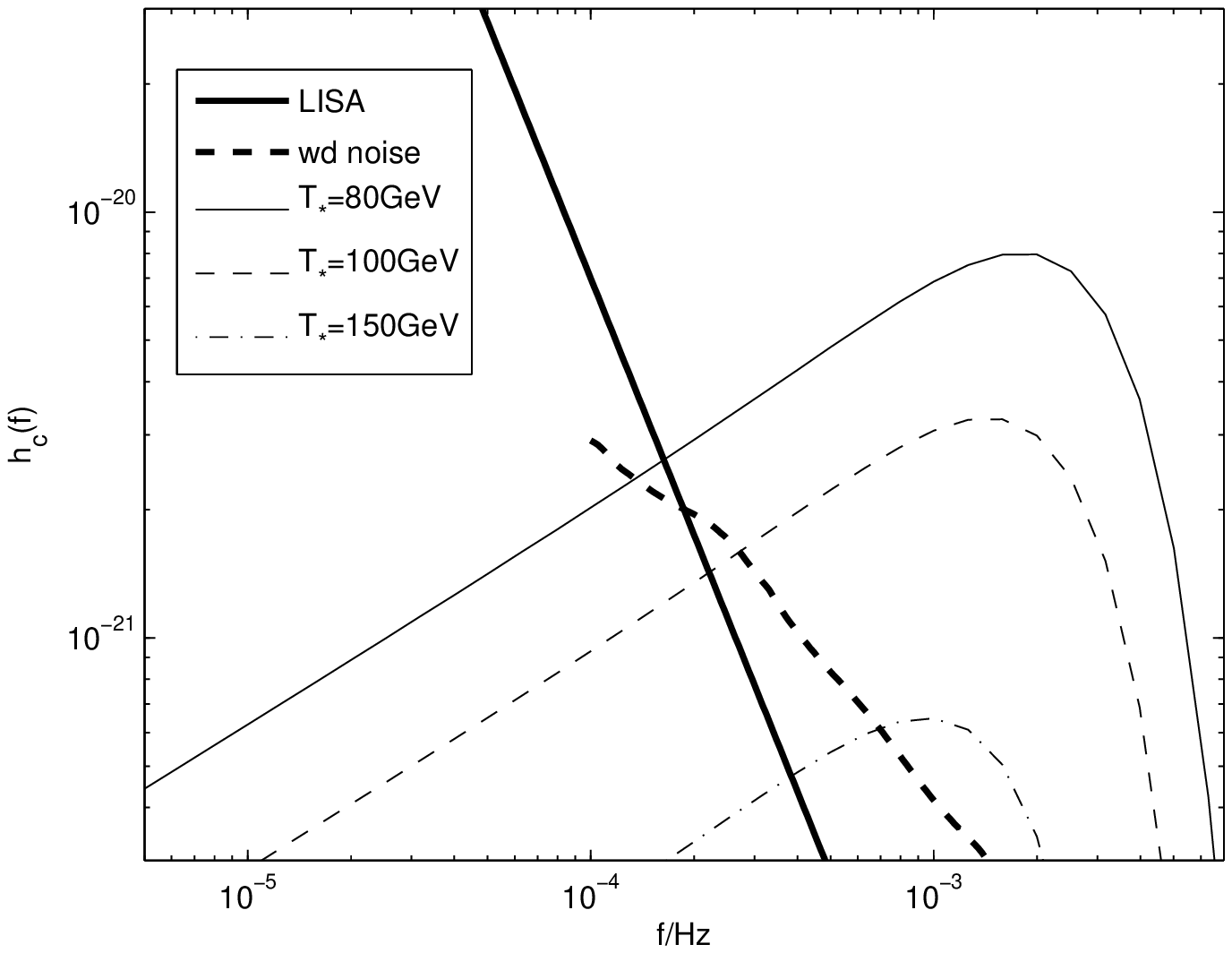,height=6cm,width=9cm}
\epsfig{file=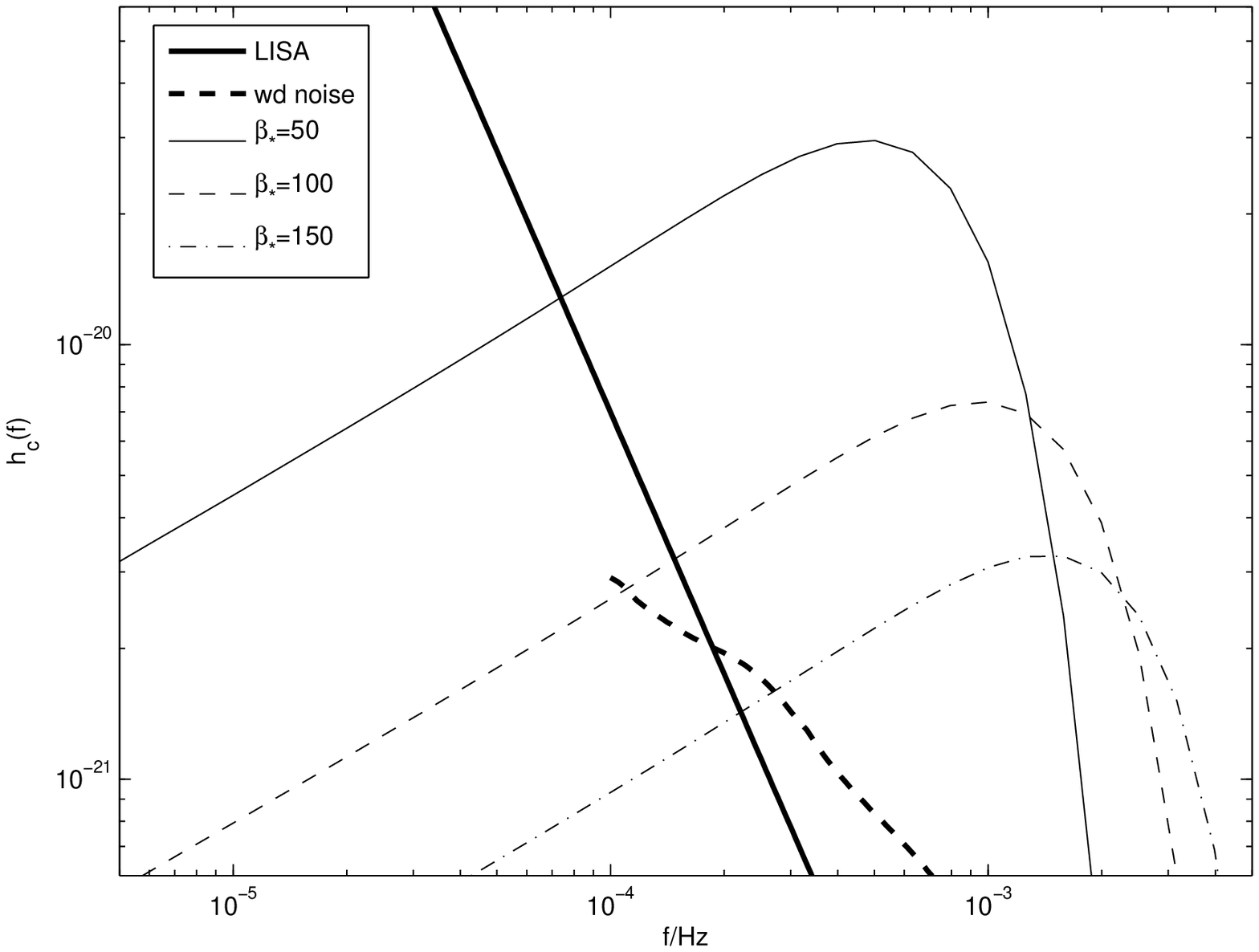,height=6cm,width=9cm} \caption{$h_C(f)$ for
the EWPT with $g_* = 100$, $v_b=1/2$, and $v_0=v_A$ with zero
magnetic helicity. The top panel: $T_* = 80$ GeV, (solid line),
$T_*= 100$ GeV (dash line), and $T_\star=150$ GeV (dash-dot line)
with $\beta=100H_\star$, while $T_\star=100$ GeV and $\beta =
50H_\star$  (solid line), $\beta=100 H_\star$ (dash line), and
$\beta=150H_\star$  (dash-dot line) in the bottom panel. In both
panels the bold solid line corresponds to the 1-year, $5\sigma$
LISA design sensitivity curve~\cite{curve} including confusion
noise from white dwarf binaries, bold dash line
~\cite{whitedwarfs}.}
\end{figure}

Eq. (\ref{hctoday}) allows us to predict the gravitational wave
spectral properties: i) The low frequency ($f\ll f_{\rm peak}$ )
dependence is $h_C(f) \propto f^{1/2}$, leading to $\Omega_{\rm
GW}(f) \propto f^3$, such a behavior is common for all causal
sources \cite{kkgm08,Caprini:2009pr} and it is true for any kind
of the waves, including the sound waves generation by turbulence
\cite{L}; ii) The peak position is determined by the
time-duration of the source, and it is equal either to $v_A \gamma
f_H$ (developed stationary source) \cite{kkgm08} or $\gamma f_H$
(pulse like source) \cite{cd,{Caprini:2006rd}}; iii) at higher
frequencies, $f\gg f_H$, the gravitational waves amplitude is
damped exponentially due to the exponential temporal
decorrelation of fluctuations \cite{kkgm08} as opposed to the
power law slow damping shape of the gravitational waves generated
by the bubble collisions
\cite{bubbles,kos1,nicolis,huber,Caprini:2009pr}; iv) for the
turbulence generated gravitational waves the power law shape
takes place in the vicinity of the peak, i.e. for $f_{\rm peak}
< f \leq {\rm R}^{1/2} f_H $, $h_C(f) \propto f^{-13/4}$
\cite{kkgm08}.

\section{Results and Discussion}
The results for $h_C(f)$ for the EWPT are shown in Fig. 1. For a
semi-analytical estimate it is straightforward to get the peak
frequency of the amplitude of the gravitational waves emitted
during direct-cascade to be \cite{kkgm08} \beq \label{f-peak}
f_{\rm peak}& =& \left(\frac{v_A}{v_b} \right) \left(
\frac{\beta}{H_\star}\right)f_H, \eeq The peak frequency is
shifted to the lower frequencies with $T_\star$ increasing due to
$f_H$ dependence on $T_\star$. Using the definition of the
$H_{ijij}(0,\omega_\star)$ tensor, Eq.~(\ref{H}), the
gravitational wave signal reaches its maximal amplitude
approximately at $f=\gamma^{-1} v_A f_H $, and then,
\begin{eqnarray}
h_C(f_{\rm peak}) & \simeq & 10^{-15} v_A^{3/2}
v_b^2 \left(\frac{\beta}{H_\star}\right)^{-2} \nonumber \\
&\times& \left(\frac{100{\rm GeV}}{T_\star}\right)
\left(\frac{100}{g_\star}\right)^{1/2}, ~ \label{h-peak}
\end{eqnarray}
According to Eqs. (\ref{f-peak}) and (\ref{h-peak}), the peak
amplitude of the EW phase transition gravitational signal is order
of $5 \times 10^{-21}$ for $T_\star =100$ GeV,
$\beta=100H_\star$, and $g_\star =100$, with $f_{\rm peak}^{\rm
EW}\simeq 10^{-3}$ Hz, which is an agreement with Fig.~1. For the
stochastic gravitational waves the real LISA sensitivity will be
lower. Even accounting for this, the gravitational signal from the
EW phase transition should be detectable, since it significantly
exceeds the LISA noise around 2mHz, if we adopt the model
described here with $v_b=1/2$, and $v_A \simeq 0.3$. Rewriting
this result in terms of the gravitational wave spectral energy
density parameter, we obtain $\Omega_{\rm GW} (f=f_{\rm peak})
\simeq 2 v_A^5 \gamma^2 (100/g_\star)^{2/3} \times 10^{-4}$. Note
$\Omega_{\rm GW}(f=f_{\rm peak})$ is the temperature independent
for a given value of $v_A$ and $v_b$, as it can be shown from the
dimensional analysis \cite{kos1}, and only slightly depends on
$g_\star$.

  The peak frequency for QCD phase transitions with $T_\star =0.15$ Gev,
$\beta=6H_\star$, and $g_\star=15$ is \beq \label{qcdpeak}
 f_{\rm peak}^{\rm QCD} &=& 1.8 \times 10^{-6}
f_{\rm peak}^{\rm EW} \simeq 2 \times 10^{-9}~{\rm Hz}. \eeq
 which is order of six magnitudes lower than the LISA low-frequency
sensitivity.

We also define the efficiency of the gravitational wave
production $\kappa_{\rm GW}$ as the ratio between the magnetic
energy density available from the phase transitions and the
energy density converted into the gravitational waves, i.e.
$\kappa_{\rm GW}(f) \equiv {\rho_{\rm GW}}/{\rho_B} $. Since both
total energy densities scale the same way with the expansion of
the Universe, $\kappa_{\rm GW}$ is invariant (no damping of the
magnetic field). It can be shown that $\kappa_{\rm GW} \simeq 2
(\beta/H_\star)^{-2} v_b^2 v_A^3 \ll 1$ \cite{kkgm08} (for EW
phase transitions with the parameters mentioned above
$\kappa_{\rm GW} \simeq 3 \gamma^2 (\rho_B/\rho_{\rm rad})^{3/2}
\simeq 10^{-6}$), so only the small fraction is transferred to
the gravitational wave signal. However, the LISA sensitivity is
$\Omega_{\rm GW}(f=1{\rm mHz}) \sim 10^{-12}$ and the magnetic
field generated signal from the EW phase transition would be
still detectable if even $1\%$ of the radiation energy consists
on the magnetic energy.

We emphasize that although we present our results for
$v_b=1/2$, they can be easily extended for an arbitrary value of
$v_b$. For example, a different model of the EW MSSM phase transition
gives $v_b$  an order of magnitude
smaller, see Ref. \cite{v}. Another degree of the freedom is
related to the energy scale of the phase transitions. In
particular, for the EW phase transition the energy
scale is approximately equal to the Higgs mass, which ranges from
about 110 GeV to 127 GeV \cite{H}, however, such freedom of the EW phase
transition energy scale just slightly affects our final results,

\section{Conclusions}

The model presented above might be applied for different
mechanisms \cite{vachaspati,pmf} of the magnetic field generation
leading to the substantial magnetic energy presence during phase
transitions, but emphasize that the ours is based on the
derivation of the magnetic field amplitude from the fundamental EW
MSSM or QCD Lagrangians.

  Summarizing, using MHD turbulence model with no helicity we
find that the gravitational wave produced during the EW phase
transitions is most likely detectable by LISA, while that
produced by the QCD phase transitions will not be detectable.
\vspace{5mm}

\acknowledgements We highly appreciate helpful discussions from L.
Samushia and useful comments from A. Brandenburg, L. Campanelli,
K. Jedamzik, A. Kosowsky, G. Lavrelashvili, B. Ratra, and A. G.
Tevzadze. T.K. acknowledges the partial support from GNSF grants
ST06/4-096, ST08/4-422, and thanks for hospitality NORDITA and the
International Center for Theoretical Physics (ICTP) where the
part of the work was performed. L. K. acknowledges support from
the NSF/INT grant 0529828.

\end{document}